\documentclass[aps,prd,onecolumn,fleqn,superscriptaddress,groupedaddress,nofootinbib,preprintnumbers]{revtex4}
\pdfoutput=1
\usepackage{amsmath,amssymb,graphicx,todonotes,array,booktabs}
\newcolumntype{L}[1]{>{\raggedright\let\newline\\\arraybackslash\hspace{0pt}}p{#1}}
\newcolumntype{C}[1]{>{\centering\let\newline\\\arraybackslash\hspace{0pt}}p{#1}}
\newcolumntype{R}[1]{>{\raggedleft\let\newline\\\arraybackslash\hspace{0pt}}p{#1}}

\newcommand{\veg}{{\sc Vegas}}

\usepackage{hyperref}
\hypersetup{
  pdfauthor={Christina Gao, Stefan Hoeche,Joshua Isaacson,Claudius Krause,Holger Schulz},
  pdftitle={Event Generation with Normalizing Flows},
  pdfkeywords={Normalizing Flows, Machine Learning, Event Generation}
}
\begin{document}
\preprint{FERMILAB-PUB-20-009-SCD-T, MCNET-20-03}
\title{Event Generation with Normalizing Flows}
\author{Christina Gao}
\affiliation{Fermi National Accelerator Laboratory,
  Batavia, IL, 60510, USA}
\author{Stefan~H{\"o}che}
\affiliation{Fermi National Accelerator Laboratory,
  Batavia, IL, 60510, USA}
\author{Joshua~Isaacson}
\affiliation{Fermi National Accelerator Laboratory,
  Batavia, IL, 60510, USA}
\author{Claudius~Krause}
\affiliation{Fermi National Accelerator Laboratory,
  Batavia, IL, 60510, USA}
\author{Holger Schulz}
\affiliation{Department of Physics,
  University of Cincinnati, Cincinnati, OH 45219, USA}
\begin{abstract}
  We present a novel integrator based on normalizing flows which can be used 
  to improve the unweighting efficiency of Monte Carlo event generators for 
  collider physics simulations. In contrast to machine learning approaches 
  based on surrogate models, our method generates the correct result 
  even if the underlying neural networks are not optimally trained.
  We exemplify the new strategy using the example of Drell-Yan type
  processes at the LHC, both at leading and partially at next-to-leading
  order QCD.
\end{abstract}
\maketitle

\section{Introduction}
Numerical simulation programs are a cornerstone of collider physics.
They are used for the planning of future experiments, analysis of current
measurements and, finally, reinterpretation based on an improved
theoretical understanding of nature. They employ Monte Carlo methods
to link theory and experiment by generating virtual collider events,
which can then be analyzed like actual events observed
in detectors~\cite{Webber:1986mc,Buckley:2011ms}.

With more and more data available from the Large Hadron Collider (LHC)
and the high-luminosity upgrade, the task of simulating collisions
at high precision becomes a matter of concern for the high-energy
physics community. The projected amount of computational resources falls
far short of the needs for precision event generation~\cite{ATLASCS}.
Past studies of the scaling behavior of multijet simulations have shown
that the computing needs are largely determined by the gradually decreasing
unweighting efficiency~\cite{Hoeche:2019rti,Buckley:2019wov}.
Except for dedicated integrators, which require a detailed understanding
of the physics problem at hand, adaptive Monte Carlo methods seem to be the only
choice to address this problem~\cite{Lepage:1977sw,Lepage:1980dq,Friedman:1981ak,
  Press:1989vk,Ohl:1998jn,Jadach:1999sf,Hahn:2004fe,Kroeninger:2014bwa}.

With the rise of machine learning, this topic has seen a resurgence of
interest recently. The possibility of using these techniques for integration
in high-energy physics was first discussed in Ref.~\cite{Bendavid:2017zhk}.
Boosted Decision Trees and Generative Adversarial Networks (GANs) were
investigated as possible general purpose integrators. This new technique improved
the integration of nonseparable high dimensional functions, for which traditional
algorithms failed. The first true physics application was presented in Ref.~\cite{Klimek:2018mza}.
The authors used Dense Neural Networks (DNN) in order to perform a variable
transformation and demonstrate that they obtain significantly larger efficiencies
for three body decay integrals than standard approaches~\cite{Alwall:2014hca}.
The major drawback of this method is its computational cost. Since the network
acts as a variable transformation, its gradient must be computed for each
inference point in order to determine the Jacobian. This becomes computationally
heavy for high multiplicity processes.

A completely orthogonal approach utilizes machine learning techniques
directly for amplitude evaluation~\cite{Bishara:2019iwh} or event generation~\cite{Otten:2019hhl,Hashemi:2019fkn,
  DiSipio:2019imz,Butter:2019cae,Carrazza:2019cnt,SHiP:2019gcl,Butter:2019eyo}.
Training data for these approaches are obtained from traditional
event generation techniques, and hence the problem of efficient
event generation still remains. In addition, one needs to ensure
that the neural networks are trained well in order to approximate the
original integrand. If this is not the case, the resulting generator will
not only be inefficient, but may actually yield the wrong result~\cite{Matchev:2020tbw}.

In this publication we propose a novel idea to address the problem:
We replace standard adaptive algorithms like \veg~\cite{Lepage:1977sw,Lepage:1980dq},
by the extension~\cite{DBLP:journals/corr/abs-1808-03856,2019arXiv190604032D}
of a Nonlinear Independent Components Estimation technique
(NICE)~\cite{DBLP:journals/corr/DinhKB14,DinhSB16}, also known
as a Normalizing Flow. This algorithm is combined with a
recursive multichannel~\cite{Kleiss:1994qy,Gleisberg:2008fv}
to form a generic integrator for collider event generation.
We test its performance in Drell-Yan type processes at the LHC,
computed both at leading and partially at next-to-leading order QCD.
We focus our study on the event generation efficiency
in comparison to the general-purpose matrix element generator
Comix~\cite{Gleisberg:2008fv}. While the training of neural networks
during the adaptation stage of the Normalizing Flow integrator
is a very time consuming operation, event generation is inexpensive,
because no gradients need to be computed. This could make the technique
a prime candidate for LHC event generation in the near future. Normalizing Flows have also been combined with Markov Chain Monte Carlo methods, showing promising results~\cite{2017arXiv170607561S,2017arXiv171109268L,2019arXiv190303704H}.

This manuscript is organized as follows. Section~\ref{sec:nice}
briefly reviews the technique of Monte Carlo integration and
introduces the concept of Normalizing Flows.
Section~\ref{sec:mlps} presents our new integrator.
Section~\ref{sec:results} discusses its computing performance and
presents some first applications to LHC event generation.
Section~\ref{sec:conclusions} contains an outlook.

\section{Normalizing Flows}
\label{sec:nice}
Monte Carlo or quasi-Monte Carlo methods are known as the only viable option
to tackle high-dimensional integration problems. The basic technique relies on
approximating the integrand by randomly sampling points in the integration
domain $\Omega$ and weighting each point, $x$, by the value of the integrand,
$f(x)$. The value of the integral is then obtained as the statistical average
of all points, and the uncertainty is determined by its variance:
\begin{equation}
  \label{eq:MC.1}
  I=\int_{\Omega} f(x)\,{\rm d}x=\frac{\Omega}{N}\sum_{i=1}^N f(x_i)=\Omega\,\langle f\rangle_{x}\;,
  \qquad
  \sigma_I=\Omega\,\sqrt{\frac{\langle f^2\rangle_{x}-\langle f\rangle_{x}^2}{N-1}}\;.
\end{equation}
In this context, $\langle\ \rangle_{x}$ indicates that the average is taken
with respect to a uniform distribution in $x$. The variance of the integral
can be reduced by importance sampling or stratified sampling~\cite{James:1968gu}.
In particular, using the transformation ${\rm d}x={\rm d}G(x)/g(x)$, with
$G(x)$ the primitive of $g(x)$, one obtains
\begin{equation}
  \label{eq:MC.3}
  I=\int_{\Omega} \frac{f(x)}{g(x)}\,{\rm d}G(x)=\Omega\,\langle f/g\rangle_{G}\;,
  \qquad
  \sigma_I=\Omega\,\sqrt{\frac{\langle (f/g)^2\rangle_{G}-\langle f/g\rangle_{G}^2}{N-1}}\;.
\end{equation}
The function $g(x)$ can now be chosen appropriately, such as to minimize the variance.
In the limit $g(x)\to f(x)/I$, Eq.~\eqref{eq:MC.3} would be estimated with vanishing uncertainty.
The goal is thus to find a distribution $g(x)$ that resembles the shape of $f(x)$ most closely,
while being integrable and invertible in order to allow for faster sampling.

For multidimensional integrals, where the variable transformation reads
${\rm d}\vec{x}\to{\rm d}\vec{x}'\,|{\rm d}\vec{x}(\vec{x}')/{\rm d}\vec{x}'|$,
we can simply replace the Jacobian $g(x)\to |{\rm d}\vec{x}'/{\rm d}\vec{x}|$,
and Eq.~\eqref{eq:MC.3} remains valid. This forms the basis for the concept of a
Normalizing Flow: For a bijective map, $G(\vec{x})$, of the random variable $\vec{x}$
that is drawn from a flat probability distribution, the variable $\vec{x}'=G(\vec{x})$
follows the probability distribution
\begin{equation}\label{eqn:qy}
  {\rm d}\vec{x}'={\rm d}G^{-1}(\vec{x}')\,\left|
  \frac{\partial G^{-1}(\vec{x}')}{\partial \vec{x}'}\right|^{-1}
  ={\rm d}\vec{x}\,\left|g(\vec{x})\right|\;.
\end{equation}
Applying a series of transformations, $G_k$, where $k=1,...,K$, one
defines the Normalizing Flow as a bijective mapping between statistical
distributions of the random variables $\vec{x}$ and $\vec{x}_K$.
\begin{equation}
  \begin{split}
    {\rm d}\vec{x}_K=&\;{\rm d}G_K(G_{K-1}(\cdots G_2(G_1(\vec{x}))))
    ={\rm d}\vec{x}\,\prod_{k=1}^{K}\,\left|
  g_k(\vec{x}_{k-1})\right|\;,
  \qquad\text{where}\qquad
  \vec{x}_0=\vec{x}\;.
  \end{split}
\end{equation}
In practice, $G_k$ is often limited to simple functions, in order to make
the determinant of the Jacobian easy to compute. This constrains the level
of complexity that can be modeled by the Normalizing Flow.
The complexity can be increased using so-called Coupling Layers,
which were first introduced in Refs.~\cite{DBLP:journals/corr/DinhKB14,DinhSB16}.
An alternative technique is based on autoregressive models~\cite{KingmaSW16,papamakarios2017masked}.
In this study we focus on Refs.~\cite{DBLP:journals/corr/abs-1808-03856,2019arXiv190604032D}
which generalizes the design of the Coupling Layers proposed by Ref.~\cite{DinhSB16}.

\subsection{Coupling Layers}
\begin{figure}[t]
  \centerline{\includegraphics[width=0.75\textwidth]{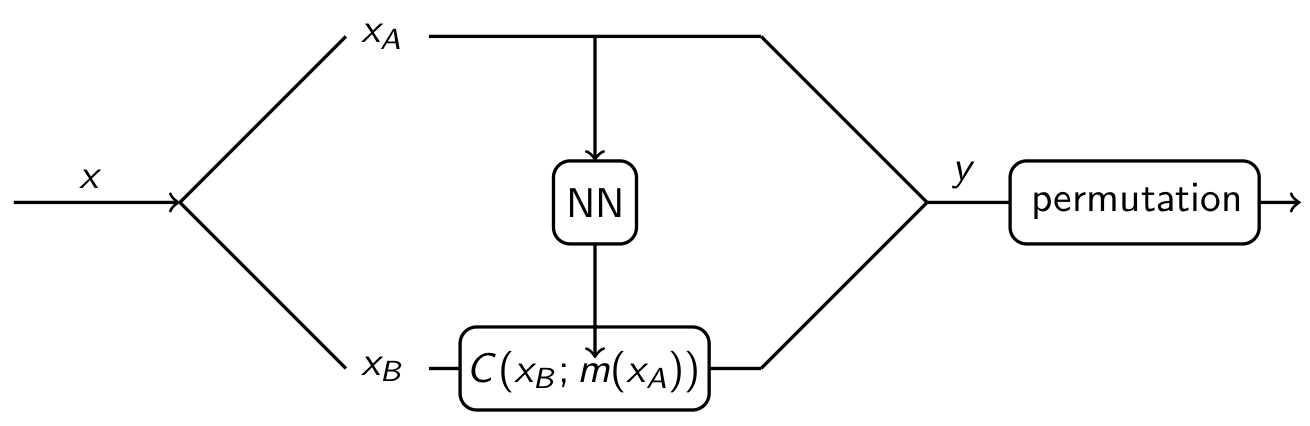}}
  \caption{Structure of a Coupling Layer.
    $m$ is the output of a neural network and defines the
    Coupling Transform, $C$, that will be applied to $x_B$.}
  \label{fig:CL}
\end{figure}
A coupling layer is a special design of a bijector, first proposed
in Refs.~\cite{DBLP:journals/corr/DinhKB14,DinhSB16}. Figure~\ref{fig:CL}
shows its basic structure. For each bijective mapping, the input
variable $\vec{x}=\{x_1,..,x_D\}$ is partitioned into two subsets,
$\vec{x}_A=\{x_1,..,x_d\}$ and $\vec{x}_B=\{x_{d+1},..,x_D\}$.
Under the bijective map, $g$, the resulting variable transforms as
\begin{equation}\label{eq:bijection}
  \begin{split}
    x_A' &= x_A, \quad\quad\quad \quad\quad\quad A\in[1,d]\;,\\
    x_B' &= C(x_B;m(\vec{x}_A)),\quad B\in[d+1,D]\;.
  \end{split}
\end{equation}
In this context, $m$ represents a neural network
that takes $x_A$ as an input and then outputs the parameters of the invertible
``Coupling Transform'', $C$, that will be applied to $x_B$.
The inverse map $G^{-1}$ is given by
\begin{equation}
  \begin{split}
    x_A &= x_A'\;,\\
    x_B &= C^{-1}(x_B';m(\vec{x}_A))=C^{-1}(x_B';m(\vec{x}_A'))\;,
  \end{split}
\end{equation}
which leads to the simple Jacobian
\begin{equation}\label{eq:jacobian_cl}
  \left|\frac{\partial G(\vec{x})}{\partial \vec{x}}\right|^{-1}=
  \left|\left(\begin{array}{cc}
    \vec{1} & 0  \\
    \frac{\partial C}{\partial m} \frac{\partial m}{\partial \vec{x}_A}
    & \frac{\partial C}{\partial \vec{x}_B}
  \end{array}\right)\right|^{-1}
  =\left|\frac{\partial C(\vec{x}_B;m(\vec{x}_A))}{\partial \vec{x}_B}\right|^{-1}\;.
\end{equation}
Note that Eq.~\eqref{eq:jacobian_cl} does not require the computation
of the gradient of $m(\vec{x}_A)$, which would scale as $\mathcal{O}(D^3)$
with $D$ the number of dimensions. In addition, since $C$ is diagonal, the computation
of the determinant of $\partial C/\partial\vec{x}_B$ scales linearly
with the number of dimensions, and is therefore tractable even for
high dimensional problems. The Normalizing Flow method is thus
evidently superior to existing integration techniques based on
Neural Networks.
To construct a complete Normalizing Flow, one simply compounds
a series of Coupling Layers with the freedom of choosing any
of the input dimensions to be transformed in each layer. 
We show in Ref.~\cite{Gao:2020vdv} that at most $2\lceil\log_2\mathrm{D}\rceil$
 Coupling Layers are required in order to express
arbitrarily complicated, nonseparable structures of the integrand.

In order to implement a Normalizing Flow integrator
in practice, the user must provide a Neural Network, represented by
$m(\vec{x}_A)$, a function $f$ to integrate, and the definition of a loss function. This is discussed in more detail in Ref.~\cite{Gao:2020vdv}.

\subsection{Piecewise Polynomial and Rational Quadratic Spline Coupling Transforms}
So far we have not yet specified the invertible coupling transforms $C$.
The design of this function impacts the flexibility of coupling layer based
Normalizing Flow algorithms and is an active field of research. A very powerful
definition of $C$ was introduced in Ref.~\cite{DBLP:journals/corr/abs-1808-03856}.
Both the domain and codomain of each Coupling Layer are defined to be the
unit hypercube. If the random variable $\vec{x}'$ is uniformly distributed,
it follows from Eq.~\eqref{eq:jacobian_cl} that the initial variable
$\vec{x}$ follows the distribution $|\partial C/\partial\vec{x}_B'|$.
Thus the coupling transform can be interpreted as the Cumulative Distribution Function
(CDF) of $\vec{x}$. Each dimension is then divided into $K$ bins and models
this CDF with a monotonically increasing polynomial function per bin. 
In particular, Ref.~\cite{DBLP:journals/corr/abs-1808-03856} experiments with
piecewise linear and piecewise quadratic coupling transforms.
In the implementation of a piecewise quadratic coupling transform,
the bin width is allowed to vary in order to increase the flexibility of
the Coupling Layer.

It may seem natural to generalize the piecewise polynomial coupling transform
to include even higher order terms in order to increase the expressivity of the
Coupling Layer. This has been proposed in Ref.~\cite{2019arXiv190604032D},
which generalized the piecewise quadratic coupling transform to allow
a monotonically increasing rational-quadratic function in each bin of
the coupling transform. To implement this, the bin heights, bin widths
and also the derivatives in between each bin are allowed to vary and are predicted by the Neural Network.

\section{Phase-Space Integration}
\label{sec:mlps}
In this section we briefly summarize the diagram-based~\cite{Byckling:1969sx}
recursive~\cite{Gleisberg:2008fv} multichannel~\cite{Kleiss:1994qy} integration 
used in our numerical routines. The latter are designed to cope with especially
large numbers of outgoing particles and exhibit exponential scaling,
reduced from factorial scaling by means of dynamic programming.

Consider a $2\to n$ scattering process, where we denote the incoming
particles by $a$ and $b$ and the outgoing particles by $1\ldots n$.
The corresponding $n$-particle differential phase space element reads
\begin{equation}
  {\rm d}\Phi_n(a,b;1,\ldots,n)=
  \left[\,\prod\limits_{i=1}^n\frac{{\rm d}^4 p_i}{(2\pi)^3}\,
    \delta(p_i^2-m_i^2)\Theta(p_{i0})\,\right]\,
    (2\pi)^4\delta^{(4)}\Big(p_a+p_b-\sum_{i=1}^n p_i\Big)\;,
\end{equation}
where $m_i$ are the on-shell masses of outgoing particles.
The full phase space can be factorized as~\cite{James:1968gu}
\begin{equation}\label{eq:split_ps}
  {\rm d}\Phi_n(a,b;1,\ldots,n)=
    {\rm d}\Phi_{n-m+1}(a,b;\pi,m+1,\ldots,n)\,\frac{{\rm d} s_\pi}{2\pi}\,
    {\rm d}\Phi_m(\pi;1,\ldots,m)\;,
\end{equation}
where $\pi=\{1,\ldots,m\}$ corresponds to a set of particle indices.
Denoting the missing subset as $\overline{\alpha}=\{a,b,1,\ldots,n\}\setminus\alpha$ 
for all $\alpha\subset \{a,b,1,\ldots,n\}$, Eq.~\eqref{eq:split_ps}
allows one to decompose the complete phase space into building blocks
corresponding to the $t$- and $s$-channel decay processes
$T_{\alpha,b}^{\,\pi,\overline{\alpha b\pi}}=\;
{\rm d}\Phi_{2}(\alpha,b;\pi,\overline{\alpha b\pi})$ 
and $S_{\pi}^{\,\rho,\pi\setminus\rho}=\;
{\rm d}\Phi_{2}(\pi;\rho,\pi\setminus\rho)$
and an $s$-channel production process $D_{\alpha,b}$,
which corresponds to overall momentum conservation
and the associated overall weight factor.
These objects have been introduced as phase space vertices
in Ref.~\cite{Gleisberg:2008fv}, while the integral 
$P_\pi={\rm d} s_\pi/2\pi$, in Eq.~\eqref{eq:split_ps},
was called a phase space propagator. In this notation,
there is a one-to-one correspondence between the computation
of hard matrix elements in the Berends-Giele recursion~\cite{
  Berends:1987me,Caravaglios:1995cd,Kanaki:2000ey,Duhr:2006iq}
and the computation of phase-space weights. Thus, the
computation of phase-space weights can be carried out
in a recursive fashion, yielding the same (exponential)
scaling as the computation of the hard matrix element.
The basic building blocks of phase space integration
can be summarized as
\begin{equation}\label{eq:ps_building_blocks}
  \begin{split}
    P_{\pi}&=\frac{{\rm d} s_\pi}{2\pi}\;,\\
    S_{\pi}^{\,\rho,\pi\setminus\rho}&=
     \frac{\lambda(s_\pi,s_\rho,s_{\pi\setminus\rho})}{
     16\pi^2\,2\,s_{\pi}}\;{\rm d}\cos\theta_\rho\,{\rm d}\phi_\rho\;,\\
    T_{\alpha,b}^{\,\pi,\overline{\alpha b\pi}}&=
     \frac{\lambda(s_{\alpha b},s_\pi,
       s_{\;\overline{\alpha b\pi}})}{16\pi^2\,2s_{\alpha b}}\;
     {\rm d}\cos\theta_\pi\,{\rm d}\phi_\pi\;,\\
    D_{\alpha,b}&=(2\pi)^4\,{\rm d}^4 p_{\,\overline{\alpha b}}\;
      \delta^{(4)}(p_\alpha+p_b
      -p_{\,\overline{\alpha b}})\;.
  \end{split}
\end{equation}
Here, $\lambda$ is given by the K{\"a}llen function
$\lambda(a,b,c)=\sqrt{(a-b-c)^2-4bc}$. Note that, in the context
of Monte Carlo integration, each basic integral ${\rm d}s$,  
${\rm d}\cos\theta$ and ${\rm d}\phi$ in Eq.~\eqref{eq:ps_building_blocks}
corresponds to a random variable, chosen in the appropriate range.
Details on the construction of the phase-space recursion
are given in Ref.~\cite{Gleisberg:2008fv}. Here we simply recall
the result in a schematic form:
\begin{itemize}
\item Select an ordered partition of the multi-index $\pi\to(\pi_1,\pi_2)$
  ($s$-channel) or $\overline{\alpha b}\to(\alpha\pi_1 b,\pi_2)$ ($t$-channel).
\item If the partition is an $s$-channel, insert an $s$-channel vertex 
  $S_{\pi}^{\,\pi_1,\pi_2}$ else insert a $t$-channel vertex
  $T_{\alpha,b}^{\,\pi_1,\pi_2}$.
\item If $\pi_1$ ($\pi_2$) is a multi-index,
  insert a propagator $P_{\pi_1}$ ($P_{\pi_2}$).
\item Proceed until there is no multi-index left.
\end{itemize}
We can improve this procedure by forming an average over all possible
ordered partitions ($\mathcal{OP}$) of each multi-index. Assigning each splitting an
adjustable weight defines the recursive multichannel. This is formalized
as follows
\begin{equation}\label{eq:ps_recursion}
  \begin{split}
    {\rm d}\Phi_S(\pi)&=
      \sum\limits_{(\pi_1,\pi_2)\in\mathcal{OP}(\pi)}\,
      \omega_\pi^{\pi_1,\pi_2}\;S_{\pi}^{\pi_1,\pi_2}\;
        P_{\pi_1}\,{\rm d}\Phi_S(\pi_1)\,
        P_{\pi_2}\,{\rm d}\Phi_S(\pi_2)
        \vphantom{{\rm d}\Phi_T^{(b)}}\;,\\
    {\rm d}\Phi_T^{(b)}(\alpha)&=
      \sum\limits_{(\pi_1,\pi_2)\in\mathcal{OP}(\overline{\alpha b})}\,
      \omega_\alpha^{\pi_1,\alpha\pi_1}\;
        T_{\alpha,b}^{\pi_1,\pi_2}\;P_{\pi_1}\,{\rm d}\Phi_S(\pi_1)\,
        P_{\pi_2}\,{\rm d}\Phi_T^{(b)}(\alpha\pi_1)+
        \omega_{\alpha,b}\;\vphantom{\Phi_T^{(b)}}\;
        D_{\alpha,b}\;{\rm d}\Phi_S(\overline{\alpha b})\;.
  \end{split}
\end{equation}
In this context we defined the one- and no-particle phase space as
${\rm d}\Phi(i)=1$ and ${\rm d}\Phi(\emptyset)=0$.
The numbers $\omega$ correspond to vertex-specific weights, which
are normalized as $\sum_{\pi_1}\omega_{\pi}^{\pi_1,\pi\setminus\pi_1}=1$
and $\sum_{\pi_1}\omega_{\alpha}^{\pi_1,\alpha\pi_1}+\omega_{\alpha,b}=1$
and can be adapted to optimize the integrator. 
The sums run over all possible $S$- and $T$-type vertices which have 
a correspondence in the matrix element. The full differential
phase space element is given by
\begin{equation}
  {\rm d}\Phi_n(a,b;1,\ldots,n)={\rm d}\Phi_T^{(b)}(a)\;.
\end{equation}
The recursive integrator is typically improved by combining it with
the \veg\ algorithm~\cite{Lepage:1977sw,Lepage:1980dq}. In this
configuration, \veg\ generates the input random numbers, $\vec{x}$,
that are used to perform the basic integrals ${\rm d}s$, 
${\rm d}\cos\theta$ and ${\rm d}\phi$ in Eq.\eqref{eq:ps_building_blocks}.
We adopt the same strategy and simply replace \veg\ by the 
Neural Network+Normalizing Flow algorithm proposed 
in Refs.~\cite{DBLP:journals/corr/DinhKB14,DinhSB16} and extended
in Refs.~\cite{DBLP:journals/corr/abs-1808-03856,2019arXiv190604032D}, which is implemented in the {\tt i-flow} package~\cite{Gao:2020vdv}.
We include the multichannel weights, $\omega$, in the integration,
which allows us to work with a single network for all channels.
All other components of the integrator remain the same.
This makes it possible to directly compare the performance to
existing algorithms. As the Normalizing Flow method is capable
of capturing correlations in the multidimensional phase space,
while still exhibiting polynomial scaling with the problem size,
one would expect a performance improvement particularly when the
number of dimensions in the problem is large.

\section{Numerical results}
\label{sec:results}
This section presents a numerical study of the novel phase-space integrator.
All computations are performed using the event generation framework
Sherpa~\cite{Gleisberg:2003xi,Gleisberg:2008ta,Bothmann:2019yzt}, 
with matrix elements provided by Comix~\cite{Gleisberg:2008fv}.
This use case differs slightly from Ref.~\cite{Bothmann:2020ywa}, where 
matrix elements are computed by Amegic~\cite{Krauss:2001iv}.
Amegic performs an explicit sum over color degrees of freedom,
while Comix uses Monte Carlo sampling. The Neural Network + Normalizing Flow
integrator could in principle be trained on a combined color-momentum space
when used with Comix. We have tested this approach but have not found 
an advantage for processes beyond $W/Z+1j$. Hence we refrain from using
this technique.

\subsection{Definition of efficiency}
\label{sec:eff_def}
In order to assess the performance of the new integrator we
investigate its unweighting efficiency. The most basic definition
of unweighting efficiency would be the average weight during
event generation (i.e.\ the integral, Eq.~\eqref{eq:MC.1}),
divided by the maximum weight.
\begin{equation}\label{eq:eff_naive}
  \eta_{\rm basic}=\frac{\langle f/g\rangle_G}{\max\{f/g\}}\;.
\end{equation}
Its inverse corresponds to the number of events to be drawn on average,
before an event is accepted with unit weight. If the distribution of weights
does not exhibit a sharp upper edge, the denominator in Eq.~\eqref{eq:eff_naive} will depend
on the sample size, and the efficiency will decrease with increasing
number of events. This is particularly worrisome when events are generated
in a distributed computing approach (e.g.\ on the LHC Computing Grid).
Each individual computing job will have its own, individual maximum,
say $\max\{f/g\}_i$. In order to realize the accuracy of the combined
event sample, each subsample must then be weighted by
$\max\{f/g\}_i/\max\limits_j\{\max\{f/g\}_j\big\}$.

Here we follow a different approach. It has recently been pointed out
that a weighted combination of event samples is prone to outliers
in the weight distribution, unless adaptive algorithms continue to be
optimized during event generation~\cite{Campbell:2019dru}.
If event generation is performed in a distributed fashion, this
cannot be achieved, as the individual compute jobs do not communicate.
The efficiency should therefore be computed based on the number of points
during the last iteration of the optimization. This definition is still
prone to possible outliers, which are removed by performing a bootstrap:
\begin{enumerate}
\item Assuming the number of events during optimization was $N_{\rm opt}$,
  draw $n N_{\rm opt}$ events.
\item From these events, select $m$ replicas of $N_{\rm opt}$ events each
  and compute their maximum weight.
\item Compute the total maximum, $w_{\rm max}$, as the median of the individual maxima.
\end{enumerate}
The efficiency is then given by
\begin{equation}\label{eq:eff_bs}
  \eta_{\rm bs}=\frac{\langle f/g\rangle_G}{w_{\rm \max}}\;.
\end{equation}
There will be a number of event weights that exceed $w_{\rm max}$.
We can account for the mismatch on an event-by-event basis by recording
their relative weight. Formally, if an event is generated with weight $w_i$,
we keep it with weight $\tilde{w}_i$, where
\begin{equation}
  \tilde{w}_i = w_{\rm max}\,\Theta\left(\frac{w_i}{w_{\rm max}}-R\right)
  \left[\frac{w_i}{w_{\rm max}}\Theta\left(\frac{w_i}{w_{\rm max}}-1\right)
    + \Theta\left(1-\frac{w_i}{w_{\rm max}}\right)\right]\;,
\end{equation}
with $R\in[0,1]$ a uniformly distributed random number.
The event sample will then be partially weighted, unweighted
against a maximum that corresponds --- on a statistical basis ---
to the largest weight probed by the adaptive integrator
during optimization. This allows one to reach in principle arbitrary
precision by applying the bootstrap and jackknife techniques
of Ref.~\cite{Efron:1992aa}. Possible practical implementations
of this method are discussed in Ref.~\cite{Campbell:2019dru}.

\subsection{Hyperparameter Optimization}
We use a quasirandom search strategy to optimize the
hyperparameters of the Neural Networks. The basic idea
of random search has been proposed in Ref.~\cite{Bergstra:2012aa},
and follows the known strategy used in tuning Monte Carlo
event generators~\cite{Buckley:2009bj}. It was pointed out
that a random search strategy typically covers the space
of hyperparameters better than a grid-based search, since
the influence of hyperparameters is often uncorrelated.
Since computing resources are limited and the computational
cost of the training is fairly large, we resort to Sobol
sequences to adequately populate the hyperparameter space,
$\mathcal{H}$.

\begin{figure}[t]
    \centering
\begin{minipage}{.48\textwidth}%
    \begin{center}
        \includegraphics[width=.98\textwidth]{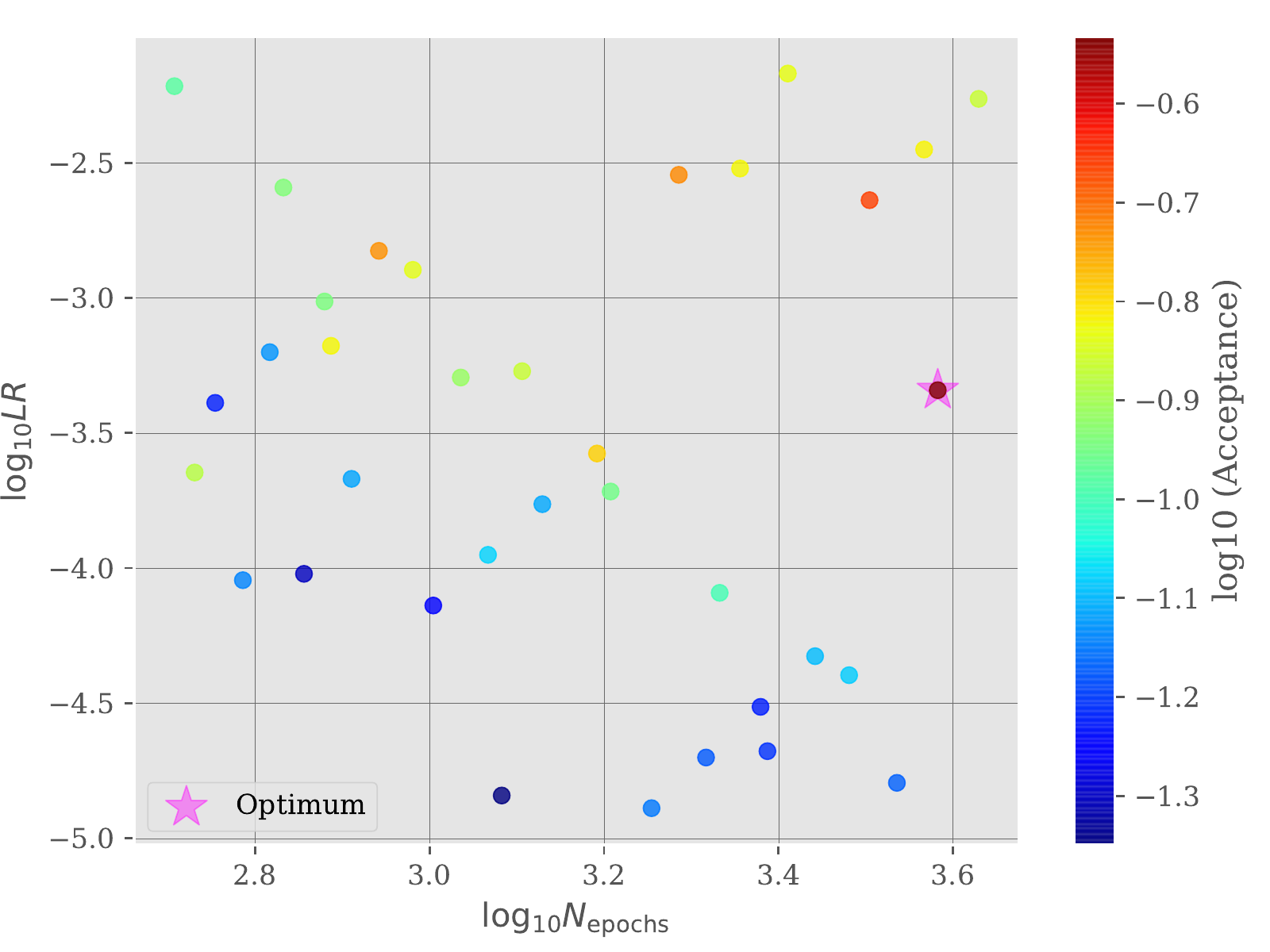}
    \end{center}
\end{minipage}%
\begin{minipage}{.04\textwidth}%
\end{minipage}%
\begin{minipage}{.48\textwidth}%
    \begin{center}
        \includegraphics[width=.98\textwidth]{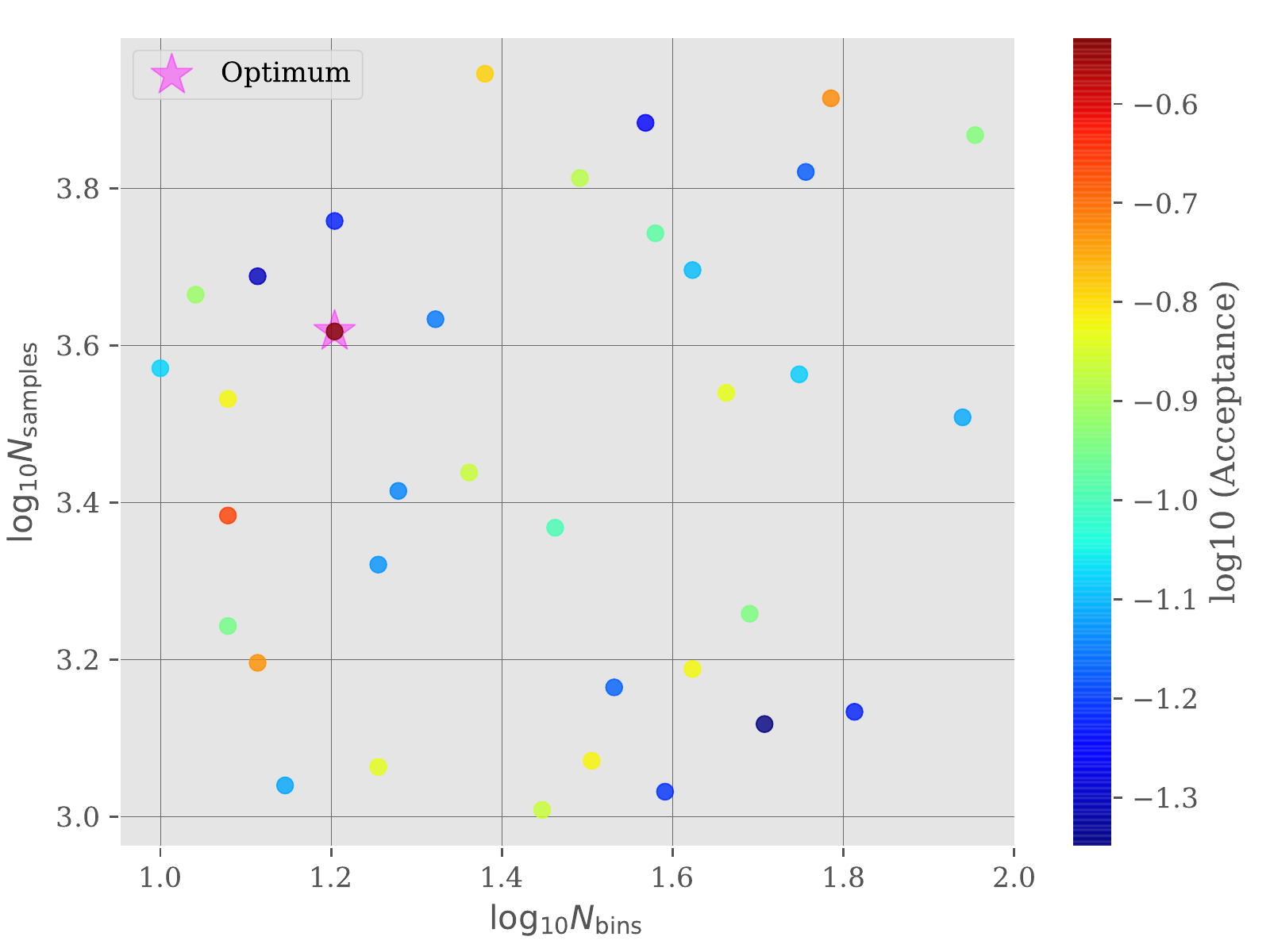}
    \end{center}
\end{minipage}%
    \caption{Projections of the sampled parameters and color coded acceptances. The plot on the left suggest a high learning rate, coupled with a large number of epochs to be beneficial. The plot on the right suggests a strong preference for a small number of bins. The best performing configuration is indicated with a star.}
    \label{fig:pscan}
    \end{figure}
\begin{table}[t]
    \centering
    \begin{tabular}{l|lll|lp{1cm}l|lll|l}
        Parameter & $c_\text{min}$   & $c_\text{max}$ & prior & $c_\text{best}$ && Parameter & $c_\text{min}$   & $c_\text{max}$ & prior & $c_\text{best}$\\ \cline{1-5}\cline{7-11}
        LR         & $10^{-5}$ & $10^{-2}$  & log & $4.55\cdot 10^{-4}$ &&
        $ N_\text{samples}$ & 1000      & 10000    & log  &  4148   \\
        LR decay rate                  & 0         & 1        & lin  &  0.534 &&
        $ N_\text{epochs}$  & 500       & 5000     & log  &  3824 \\
        LR decay $e$-folds            & 2         & 10       & lin  &  5      &&
        $ N_\text{bins}$    & 10     & 100       & log  &  16 \\
        $N_\text{nodes}^\text{max}$  & $2^5$     & $2^9$    & lin & $2^9$ &&
        $N_\text{layers}$            & 6         & 10       & lin & 8 \\
    \end{tabular}
    \caption{Parameter ranges and prior distributions in $\mathcal{H}$ 
    and values of the optimal point in the scan. We use LR as abbreviation for the learning rate.}
\label{tab:sampling boundaries}
\end{table}

The figure of merit for a configuration $c\in\mathcal{H}$ is
its unweighting efficiency. Due to the limited computing resources
we redefine for the purpose of tuning $w_{\rm max}$ in Eq.~\eqref{eq:eff_bs}
to be the 99.9th percentile of the distribution of nonzero absolute values
of weights. We have tested that this makes the definition
of $w_{\rm max}$ stable against variations in the random
seed and gives results that are qualitatively compatible
with the technique described in Sec.~\ref{sec:eff_def}.
The main advantage of the alternative definition is that it 
allows us to determine the target number of points for the computation
of $w_{\max}$ as $1/(\eta_{\rm cut}\sigma^2\varepsilon)$,
where $\eta_{\rm cut}$ is the cut efficiency, $\sigma$
is the desired Monte Carlo accuracy, and $\varepsilon$
is the desired percentile defining $w_{\rm max}$.

Table~\ref{tab:sampling boundaries} shows the sampling boundaries for $\mathcal{H}$, and the optimal choice for $W+1$~jet. Figure~\ref{fig:pscan} shows the distribution of sampling points in the $N_{\rm epochs}$-$LR$ and $N_{\rm bins}$-$N_{\rm samples}$
planes. The first suggests a high learning rate, coupled with a large number of epochs to be beneficial. The second suggests a strong preference for a small number of bins.
We also performed a parameter scan for $W+2$~jets and found the best configuration
to be comparable to the above and to yield similar unweighting
efficiency as the best configuration in the $W+1$~jet setup.
Due to limited computing resources, we did not perform a separate hyperparameter
scan for $W+\ge3$~jets.

\subsection{Comparison to existing approaches}
\begin{table*}[t]
  \centering
  \begin{tabular}{l@{\hspace*{3mm}}l|C{15mm}C{15mm}C{15mm}C{15mm}C{15mm}|C{15mm}C{15mm}}
    \multicolumn{2}{l|}{unweighting efficiency} & \multicolumn{5}{c|}{LO QCD}
    & \multicolumn{2}{c}{NLO QCD (RS)} \\[1mm]
    \multicolumn{2}{l|}{$\langle w\rangle/w_{\rm max}$}
    & $n=$0 & $n=$1 & $n=$2 & $n=$3 & $n=$4 & $n=$0 & $n=$1 \\[1mm]\hline
    $W^++n\;{\rm jets}$
    & Sherpa \vphantom{$\int_A^{B^C}$} & $2.8\cdot10^{-1}$ & $3.8\cdot10^{-2}$ & $7.5\cdot10^{-3}$ & $1.5\cdot10^{-3}$ & $8.3\cdot10^{-4}$ & $9.5\cdot10^{-2}$ & $4.5\cdot10^{-3}$ \\
    & NN+NF \vphantom{$\int_A^{B^C}$} & $6.1\cdot10^{-1}$ & $1.2\cdot10^{-1}$ & $1.0\cdot10^{-2}$ & $1.8\cdot10^{-3}$ & $8.9\cdot10^{-4}$ & $1.6\cdot10^{-1}$ & $4.1\cdot10^{-3}$ \\
    & Gain\vphantom{$\int_A^{B^C}$} & 2.2 & 3.3 & 1.4 & 1.2 & 1.1 & 1.6 & 0.91 \\\hline
    $W^-+n\;{\rm jets}$
    & Sherpa \vphantom{$\int_A^{B^C}$} & $2.9\cdot10^{-1}$ & $4.0\cdot10^{-2}$ & $7.7\cdot10^{-3}$ & $2.0\cdot10^{-3}$ & $9.7\cdot10^{-4}$ & $1.0\cdot10^{-1}$ & $4.5\cdot10^{-3}$ \\
    & NN+NF \vphantom{$\int_A^{B^C}$} & $7.0\cdot10^{-1}$ & $1.5\cdot10^{-1}$ & $1.1\cdot10^{-2}$ & $2.2\cdot10^{-3}$ & $7.9\cdot10^{-4}$ & $1.5\cdot10^{-1}$ & $4.2\cdot10^{-3}$ \\
    & Gain\vphantom{$\int_A^{B^C}$} & 2.4 & 3.3 & 1.4 & 1.1 & 0.82 & 1.5 & 0.91 \\\hline
    $Z+n\;{\rm jets}$
    & Sherpa \vphantom{$\int_A^{B^C}$} & $3.1\cdot10^{-1}$ & $3.6\cdot10^{-2}$ & $1.5\cdot10^{-2}$ & $4.7\cdot10^{-3}$ & & $1.2\cdot10^{-1}$ & $5.3\cdot10^{-3}$ \\
    & NN+NF \vphantom{$\int_A^{B^C}$} & $3.8\cdot10^{-1}$ & $1.0\cdot10^{-1}$ & $1.4\cdot10^{-2}$ & $2.4\cdot10^{-3}$ & & $1.8\cdot10^{-3}$ & $5.7\cdot10^{-3}$ \\
    & Gain\vphantom{$\int_A^{B^C}$} & 1.2 & 2.9 & 0.91 & 0.51 & & 1.5 & 1.1 \\
    \end{tabular}
    \caption{Unweighting efficiencies at the LHC at $\sqrt{s}=14~{\rm TeV}$
      using the NNPDF 3.0 NNLO PDF set and a correspondingly defined strong coupling.
      Jets are identified using the $k_T$ clustering algorithm
      with $R=0.4$, $p_{T,j}>20\;{\rm GeV}$ and $|\eta_j|<6$.
      In the case of $Z/\gamma^*$ production, we also apply
      the invariant mass cut $66<m_{ll}<116{\rm GeV}$.}
    \label{tab:eff_comparison}
\end{table*}

In this section we compare the performance of our integrator based on the 
normalizing flow technique to the best alternative method available in the
public event generation program Sherpa~\cite{
  Gleisberg:2003xi,Gleisberg:2008ta,Bothmann:2019yzt}.
The basic computational setup is analogous to Ref.~\cite{Hoeche:2019rti}.
We consider $W^\pm$- and $Z$-boson production in proton-proton collisions
at the high-luminosity LHC at $\sqrt{s}=14~{\rm TeV}$. We use the NNPDF3.0 NNLO
PDF set~\cite{Ball:2017nwa} and evaluate the strong coupling accordingly.
Jets are defined using the $k_T$ clustering algorithm with $R=0.4$,
$p_{T,j}>20\;{\rm GeV}$ and $|\eta_j|<6$. Following the good agreement
between parton-level and particle-level results established
in Ref.~\cite{Bellm:2019yyh}, and the good agreement between fixed-order
and MINLO~\cite{Hamilton:2012np} results established in Ref.~\cite{Anger:2017nkq},
the renormalization and factorization scales are set to $\hat{H}_T'/2$~\cite{Berger:2010zx}.

Table~\ref{tab:eff_comparison} shows a comparison of unweighting efficiencies
defined according to Eq.~\eqref{eq:eff_bs}, where $N_{\rm opt}=2\cdot 10^4$, $n=50$, and $m=10^3$.
We give results for both leading-order cross sections, and for the subtracted
real-emission corrections to next-to-leading order cross sections, using the dipole
method of Catani and Seymour~\cite{Catani:1996vz}. The subtracted real-emission
corrections typically present the biggest challenge in cross-section calculations
at next-to-leading order in the perturbative expansion, and therefore drive 
the computing demands of precision simulations for LHC experiments.
The new integrator based on Neural Networks and Normalizing Flows gives 
a much larger unweighting efficiency than Sherpa 
in processes  with few jets, both at LO and at NLO precision. 
In processes with more final-state jets it performs
similarly to the existing integration techniques in Sherpa.
The Neural Network technique generally performs better when we do not 
combine it with a multichannel approach for initial-state integration. 
Only the major features of the final state should be mapped out, 
for example the Breit-Wigner resonance in $W^\pm$ production.
This indicates, unsurprisingly, that the Normalizing Flow approach 
is more efficient in approximating smooth structures of the integrand
than in differentiating between effectively independent integration domains.
It leads to the strikingly lower efficiencies in $Z$-boson production processes,
where we have combined a $1/\hat{s}$ integrator and a Breit-Wigner mapping.
In general, the new method is best applied to low-multiplicity problems,
where the training of the neural networks can be performed at reasonable speed
with relatively few samples per epoch. We expect that in the high multiplicity
cases the Neural Network + Normalizing Flow technique will also outperform Sherpa,
if it can be trained over sufficiently many epochs with sufficiently many 
sample points. However, due to restricted computing resources, we were not able
to verify this claim for the case of $W^\pm/Z$+4j production. The picture might be
altered by future implementations of matrix element generators on accelerators.
For exploratory work on this topic, see Refs.~\cite{Giele:2010ks,Hagiwara:2013oka}.

\section{Conclusions}
\label{sec:conclusions}
We have presented a novel approach to phase-space integration for collider
physics simulations, which is based on Neural Networks and Normalizing Flows. 
The integrator is implemented as an add-on to the existing event generator
Sherpa and can be used with both internal matrix-element generators,
Comix and Amegic. Neural Network hyperparameters were tuned using a 
quasirandom search strategy. For the optimal set of parameters, 
the unweighting efficiency of the integrator exceeds that of conventional
methods by a factor 2--3 in simple processes.
In high-multiplicity processes, traditional techniques tend to perform
similarly well, while also requiring fewer computing resources. 
We expect this picture to change as implementations of matrix element 
generators on accelerators such as GPUs and TPUs become available.
Additional possible improvements of Normalizing Flows are discussed in Ref.~\cite{Gao:2020vdv}. These findings are corroborated by the results presented in Refs.~\cite{Bothmann:2020ywa,Gao:2020vdv}.

\section*{Acknowledgments}
We are grateful to John Campbell and Tobias Neumann for many
interesting discussions on the definition of unweighting efficiency.
We thank Enrico Bothmann, Marek Sch{\"o}nherr, Steffen Schumann and
Frank Siegert for comments and for discussions on unweighted
event generation and adaptive integration using Neural Networks.
We thank Tilman Plehn for discussions on Neural Networks and
Joao M. Goncalves Caldeira, Felix Kling, Luisa Lucie-Smith, Nhan Tran,
and the participants of the Aspen workshop ``The Energy Frontier
Beyond the LHC Run 2" for their comments.

This manuscript has been authored by Fermi Research Alliance,
LLC under Contract No.\ DE-AC02-07CH11359 with the U.S.\ Department
of Energy, Office of Science, Office of High Energy Physics.
The work of H.S.\ and S.H.\ was supported by the U.S. Department of Energy,
Office of Science, Office of Advanced Scientific Computing Research,
Scientific Discovery through Advanced Computing (SciDAC) program,
grants ``HEP Data Analytics on HPC'', No.~1013935 and
``HPC framework for event generation at colliders.''
C.K.\ acknowledges the support of the Alexander von Humboldt Foundation.
This work was performed in part at Aspen Center for Physics,
which is supported by National Science Foundation Grant No. PHY-1607611.
It used computing resources of SLAC National Accelerator Laboratory,
and of the National Energy Research Scientific Computing Center (NERSC),
a U.S. Department of Energy Office of Science User Facility
operated under Contract No.\ DE-AC02-05CH11231.
\bibliography{ML_PT}

\end{document}